\documentstyle[12pt]{article}

\begin{document}
\noindent {\large \bf B\"acklund transformations for
nonlinear evolution equations: Hilbert space approach}\\

\vspace{1.5cm}
\noindent Krzysztof Kowalski\\

\vspace{1.5cm}
\noindent Department of Biophysics, Institute of Physiology and
Biochemistry, Medical School of \L\'od\'z, 3 Lindley St.,
90-131 \L\'od\'z, Poland
\vspace{1.5cm}
\begin{abstract}
A new method of determining B\"acklund transformations for
nonlinear partial differential equations of the evolution
type is introduced. Using the Hilbert space approach the
problem of finding B\"acklund transformations is brought
down to the solution of an abstract equation in Hilbert
space.
\end{abstract}
\vspace{1.5cm}

\vspace{1.5cm}
\newpage
\noindent {\bf 1.~Introduction}\vspace{.8cm}

In recent years, the interest in the B\"acklund
transformations is steadily increasing in connection with a
general increase in the understanding of methods for
solution of nonlinear partial differential equations. Let us
only recall the r\^ole played by the Miura transformation in
discovery of the method of inverse scattering [1].

In the recent paper [2] we introduced a new method for
finding linearization transformations for nonlinear partial
differential equations of the evolution type based on the
Hilbert space approach to the theory of nonlinear dynamical
systems developed by the author [2-7]. The theory was
illustrated by an example of the Burgers equation (we
obtained in a simple way the celebrated Hopf-Cole
transformation). The present work is devoted to a
generalization of the treatment to the case involving
general B\"acklund transformations. Proceeding analogously
as in the paper [2] we demonstrate that the problem of
finding B\"acklund transformations for evolution equations
can be reduced to the solution of an abstract equation in
Hilbert space. We illustrate the algorithm on the example of
the Miura transformation.\vspace{1.2cm}

\noindent {\bf 2.~B\"acklund transformations}\vspace{.8cm}

We begin with a brief account of the Hilbert space
description of nonlinear partial differential equations of
the evolution type [4]. Consider the equation \vspace{.4cm}
\begin{equation}
{\partial_t}u(x,t) = F(u,{D^\alpha}u),\qquad u(x,0)={u_0}(x),
\end{equation}

\vspace{.4cm}
\noindent where $u: {\bf R}^s\times{\bf R}\to {\bf R},\:
{D^\beta}={\partial^{|\beta|}}/\partial x_1^{\beta_1}\ldots
\partial
x_s^{\beta_s},\:|\beta|=\sum\limits_{i=1}^{s}\beta_i,\: F$ is
analytic in $u$,\\ ${D^\alpha}u$ and ${u_0}\in {L_{\bf R}^2}({{\bf
R}^s},{d^s}x)$ (real Hilbert space of square-integrable
functions).
\newpage
\noindent Let $|u\rangle$ be a normalized functional
coherent state (see appendix), where $u$ satisfies (1) and
we assume that $u$ is square-integrable. We define the
vectors $|u,t\rangle$ as follows \vspace{.4cm}
\begin{equation}
|u,t\rangle = \exp\left[\frac{1}{2}\left(\int {d^s}xu^2
-\int {d^s}xu_0^2\right)\right]|u\rangle.
\end{equation}

\vspace{.4cm}
\noindent Suppose now we are given a boson operator of the
form
\vspace{.4cm}
\begin{equation}
M = \int {d^s}x{a^\dagger}(x)F(a(x),{D^\alpha}a(x)),
\end{equation}

\vspace{.4cm}
\noindent where ${a^\dagger}(x)$ and $a(x)$ are the standard
Bose field operators.

\vspace{.8cm}
\noindent An easy differentiation shows that the vectors (2)
satisfy the following linear evolution equation in Hilbert
space
\vspace{.4cm}
\begin{equation}
\frac{d}{dt}|u,t\rangle = M|u,t\rangle,\qquad
|u,0\rangle=|u_0\rangle.
\end{equation}

\vspace{.4cm}
\noindent Taking into account (2) we find that the following
eigenvalue equation holds true
\vspace{.4cm}
\begin{equation}
a(x)|{u_0},t\rangle=u[{u_0}|x,t]|{u_0},t\rangle,
\end{equation}

\vspace{.4cm}
\noindent where $|{u_0},t\rangle$ is the solution of (4) and
$u[{u_0}|x,t]$ is the solution of (1) (the square brackets
designate the functional dependence of $u$ on $u_0$).

\vspace{.8cm}
\noindent It thus appears that the nonlinear equation (1)
can be brought down to the linear abstract
Schr\"odinger-like equation (4). The restriction to square
integrable data is not too serious. Indeed, the approach was
shown to work also in the case when the initial data were
not square integrable [2,4]. The postulate that the
solutions are square integrable at any time is rather
restrictive one. It should be noted, however, that there
exist numerous equations of classical and of current
interest satisfying this requirement. Examples include
Korteweg-de Vries equation, Burgers equation, nonlinear
Schr\"odinger equation and Kadomtsev-Petviashvili equation.
We should also mention that the treatment can be immediately
extended to the case of complex multidimensional systems of
partial differential equations (1) with a right-hand side
dependent on $x, t$ [4].

We now discuss the transformation of variables within the
Hilbert space approach. Consider the following
transformation
\vspace{.4cm}
\begin{equation}
u^\prime = \phi[u|x],
\end{equation}

\vspace{.4cm}
\noindent where $\phi$ is analytic in $u$.

\vspace{.8cm}
\noindent Taking into account (2) we find that under (6) the
``Hamiltonian'' $M$ transforms as

\vspace{.4cm}
\begin{equation}
M^\prime = \int{d^s}x{a^\dagger}(x)[\phi[a|x],M].
\end{equation}

\vspace{.4cm}
\noindent Therefore, whenever the transformation (6)
converts the equation (1) into the equation
\vspace{.4cm}
\begin{equation}
{\partial
_t}{u^\prime}={F^\prime}({u^\prime},{D^\beta}u^\prime)
\end{equation}

\vspace{.4cm}
\noindent then the following commutation relation holds
\vspace{.4cm}
\begin{equation}
[\phi[a|x],M]={F^\prime}(\phi[a|x],{D^\beta}\phi[a|x]).
\end{equation}

\vspace{.4cm}
\noindent On taking the Hermitian conjugate of (9) and using (B.7) we
arrive at the following equation
\vspace{.4cm}
\begin{equation}
{M^\dagger}|\phi(x)\rangle={F^\prime}(\phi[{a^\dagger}|x],{D
^\beta}\phi[{a^\dagger}|x])|0\rangle,
\end{equation}

\vspace{.4cm}
\noindent where
$|\phi(x)\rangle=\phi[{a^\dagger}|x]|0\rangle.$

\vspace{.8cm}
\noindent The vector $|\phi(x)\rangle$ is related to the
B\"acklund transformation (6) by
\vspace{.4cm}
\begin{equation}
\phi[u|x]=\langle
u|\phi(x)\rangle\exp\left(\frac{1}{2}\int{d^
s}xu^2\right).
\end{equation}

\vspace{.4cm}
\noindent It thus appears that the problem of determining
B\"acklund transformation (6) is equivalent to solving
Hilbert space equation (10). The particular case
\vspace{.4cm}
\begin{equation}
{F^\prime}({u^\prime},{D^\beta}u^\prime)=Lu^\prime,
\end{equation}

\vspace{.4cm}
\noindent where $L$ is a linear differential operator,

\vspace{.8cm}
\noindent when (6) is the linearization transformation and
(10) takes the form
\vspace{.4cm}
\begin{equation}
{M^\dagger}|\phi(x)\rangle=L|\phi(x)\rangle
\end{equation}

\vspace{.4cm}
\noindent was discussed in ref.~2. Solving (13) we obtained
in a simple way the celebrated Hopf-Cole transformation
reducing the Burgers equation to the heat equation. We note
that the case of the linearization transformation is the
only one when (10) is linear. Nevertheless, it appears that
there exist nontrivial cases when the B\"acklund
transformations can be determined easily by solving (10).
We now illustrate this observation by the example of the
Miura transformation.
\newpage
\noindent{\it Example}.\quad Consider the modified
Korteweg-de Vries
equation
\vspace{.4cm}
\begin{equation}
{\partial_t}u = -{\partial_x^3}u + 6u^2{\partial_x}u
\end{equation}

\vspace{.4cm}
\noindent and the Korteweg-de Vries equation
\vspace{.4cm}
\begin{equation}
{\partial_t}u^\prime = -{\partial_x^3}u^\prime +
6u^\prime{\partial_x}u^\prime.
\end{equation}

\vspace{.4cm}
\noindent We seek for the B\"acklund transformation $\phi$ such that
\vspace{.4cm}
\begin{equation}
u^\prime=\phi[u|x].
\end{equation}

\vspace{.4cm}
\noindent Eq.~[10] corresponding to (16) can be written as
\vspace{.4cm}
\begin{equation}
M^\dagger|\phi(x)\rangle = -\partial_x^3|\phi(x)\rangle  +
 6\phi[a^\dagger|x]\partial_x|\phi(x)\rangle ,
\end{equation}

\vspace{.4cm}
\noindent where the conjugation $M^\dagger$ of the ``Hamiltonian''
corresponding to (14) is
\vspace{.4cm}
\begin{equation}
M^\dagger = \int dx(-a^{\dagger\prime\prime\prime}(x) +
6a^{\dagger2}(x)a^{\dagger\prime}(x))a(x).
\end{equation}

\vspace{.4cm}
\noindent On writing (18) in the coordinate representation (see
appendix A) we obtain
\vspace{.4cm}
\setcounter{equation}{19}
$$\partial_{x_1}^3\phi_1(x;x_1)=
-\partial_x^3\phi_1(x;x_1),\eqno (\theequation \rm a)$$
$$(\partial_{x_1}^3 +
\partial_{x_2}^3)\phi_2(x;x_1,x_2)=-\partial_x^3\phi_2(x;x_1,x_2
)$$
$$\null + 6\phi_1(x;x_2)\partial_x\phi_1(x;x_1) +
6\phi_1(x;x_1)\partial_x\phi_1(x;x_2),\eqno (\theequation \rm
b)$$
\newpage
$$\sum\limits_{i=1}^{n+2}\partial_{x_i}^3 \phi_{n+2}(x;x_1,\ldots
,x_{n+2})$$
$$\null -12\sum\limits_{i=1}^{n+2}\sum\limits_{r,s\ne i
\atop r>
s}\partial_{x_i}[\delta(x_i-x_r)\delta(x_i-x_s)\phi_n(x;x_1,
\ldots ,\check x_r,\ldots ,\check x_s,\ldots ,x_{n+2})]$$
$$=-\partial_x^3\phi_{n+2}(x;x_1,\ldots ,x_{n+2})$$
$$\null+6\sum\limits_{r=1}^{n+1}\frac{1}{r!}\sum\limits_{i_
1=1}^{n+2}\sum\limits_{i_2\ne i_1}\ldots\sum\limits_{i_r\ne
i_{r-1}}\phi_{n+2-r}(x;x_1,\ldots ,\check x_{i_1},\ldots
,\check x_{i_r},\ldots ,x_{n+2})$$
$$\times \partial_x\phi_r(x;x_{i_1},\ldots
,x_{i_r}),\qquad n=1,2,\ldots ,\infty, \eqno (\theequation
\rm c)$$

\vspace{.4cm}
\noindent where $\phi_n(x;x_1,\ldots ,x_n)=\langle
x_1,\ldots ,x_n|\phi(x)\rangle $ and the reversed hat
over $x_r,\:x_s$ and $x_{i_1},\:x_{i_r}$ denotes that these
variables should be omitted from the set $\{x_1,\ldots
,x_{n+2}\}$.

\vspace{.8cm}
\noindent Hence, passing to the Fourier transformation we
get
\setcounter{equation}{20}
\vspace{.4cm}
$$(k^3+k_1^3)\tilde \phi_1(k;k_1) = 0,\eqno (\theequation \rm a)$$
$$(k^3+k_1^3+k_2^3)\tilde \phi_2(k;k_1,k_2)=-6k\int
dk'\tilde \phi_1(k-k';k_1)\tilde
\phi_1(k',k_2),\eqno(\theequation\rm b)$$
$$\left(k^3+\sum\limits_{i=1}^{n+2}k_i^3\right)\tilde
\phi_{n+2}(k;k_1,\ldots ,k_{n+2})$$
$$\null+12\sum\limits_{i=1}^{n+2}\sum\limits_{r,s\ne i\atop
r>s}k_i\tilde \phi_n(k;k_1,\ldots ,\check k_r,\ldots ,\check
k_s,\ldots ,k_{n+2})_{|k_i\to k_i+k_r+k_s}$$
$$=-6\sum\limits_{r=1}^{n+1}\frac{1}{r!}\sum\limits_{i_1=1}^{
n+2}\sum\limits_{i_2\ne i_1}\ldots \sum\limits_{i_r\ne
i_{r-1}}\int dk'\tilde \phi_{n+2-r}(k-k';k_1,\ldots ,\check
k_{i_1},\ldots ,\check k_{i_r},\ldots ,k_{n+2})$$
$$\times k'\tilde \phi_r(k';k_{i_1},\ldots ,k_{i_r}),\qquad
n=1,2,\ldots ,\infty. \eqno (\theequation\rm c)$$
\newpage
\noindent Making use of the identities
\setcounter{equation}{21}
\vspace{.4cm}
$$(k^3+k_1^3)\delta(k+k_1)=0, \eqno (\theequation\rm a)$$
$$\sum\limits_{i=0}^{n}k_i^3\delta\left(\sum\limits_{i=0}^{n
}k_i\right)
=3\sum\limits_{q>r>s}k_qk_rk_s\delta\left(\sum\limits_
{i=0}^{n}k_i\right),\eqno (\theequation\rm b) $$

\vspace{.4cm}
\noindent where $q,r,s\in\{0,1,\ldots ,n\},\:n\ge2$ and we set
$k_0 = k$,
one finds easily the following solution to (20)
\vspace{.4cm}
\setcounter{equation}{22}
$$\tilde \phi_1=-ik\delta(k+k_1),\qquad \tilde
\phi_2=2\delta(k+k_1+k_2),\qquad \tilde \phi_n=0, \qquad
n\ge3.\eqno (\theequation)$$

\vspace{.4cm}
\noindent On performing Fourier's inverse transformations to
(22) and using
\vspace{.4cm}
\begin{equation}
|\phi(x)\rangle =\sum\limits_{n=1}^{\infty}\frac{1}{n!}\int
dx_1\ldots dx_n\phi_n(x;x_1,\ldots ,x_n)|x_1,\ldots
,x_n\rangle ,
\end{equation}

\vspace{.4cm}
\noindent we obtain the solution to (17) of the form
\vspace{.4cm}
\begin{equation}
|\phi(x)\rangle =|xx\rangle + \partial_x|x\rangle .
\end{equation}

\vspace{.4cm}
\noindent Hence taking into account (11) and (B.5) we finally arrive
at the desired B\"acklund transformation (16) such that
\vspace{.4cm}
\begin{equation}
u'=u^2+\partial_xu.
\end{equation}

\vspace{.4cm}
\noindent The mapping (25) coincides with the celebrated
Miura transformation relating solutions of the modified
Korteweg-de Vries equation to solutions of the Korteweg-de
Vries equation.
\newpage
\noindent {\bf Conclusion}

\vspace{.8cm}
Applying the Hilbert space appproach to the theory of
nonlinear dynamical systems developed by the author a new
method is introduced in this work of finding B\"acklund
transformations for nonlinear evolution equations. It should
be noted that regardless of the form of eqs.~(20b) and (20c)
we have rederived the Miura transformation from (20) in
purely algebraic mannner (we need not have solved any
integral equation). The algorithm described herein is an
example of the following general technique of the study of
nonlinear partial differential equations based on the
Hilbert space formalism. Namely, using the Hilbert space
approach we first derive an abstract equation in Hilbert
space corresponding to the considered nonlinear evolution
problem. Then writing this equation in the coordinate
representation and performing a Fourier transformation we
obtain a system of algebraic equations related to the
original problem. This technique was succesfully applied for
finding first integrals [4,6] and linearization
transformations [2] for nonlinear partial differential
equations of the evolution type. The simplicity of the
algorithm for determining B\"acklund transformations
described in this work suggests that it would also be a
useful tool in the study of nonlinear evolution equations.

\vspace{1.2cm}
\noindent {\bf Acknowledgements}

\vspace{.8cm}
\noindent This work was supported by KBN grant 2 0903 91 01.
\newpage
\noindent {\bf Appendix A.~Coordinate representation}

\vspace{.8cm}
We first recall the basic properties of the coordinate
representation. The Bose creation ($a^\dagger(x)$) and
annihilation ($a(x)$) operators obey the canonical
commutation relations
\vspace{.4cm}
\setcounter{equation}{1}
$$[a(x),a^\dagger(x')] = \delta(x-x'),$$
\hfill (A.1)
$$[a(x),a(x')]=[a^\dagger(x),a^\dagger(x')]=0,\qquad
x,x'\in{\bf R}^s.$$

\vspace{.4cm}
\noindent Let us assume that there exists in a Hilbert space
of states $\cal H$ where act Bose operators a unique
normalized vector $|0\rangle $ (vacuum vector) such that
\vspace{.4cm}
\setcounter{equation}{2}
$$a(x)|0\rangle =0,\qquad \hbox{for every}\ x\in{\bf R}^s.
\eqno (\rm A.\theequation)$$

\vspace{.4cm}
\noindent We also assume that there is no nontrivial closed
subspace of $\cal H$ which is invariant under the action of
the operators $a(x),a^\dagger(x')$. The state vectors
defined as
\vspace{.4cm}
\setcounter{equation}{3}
$$|x_1,\ldots ,x_n\rangle =
\left(\prod\limits_{i=1}^{n}a^\dagger(x_i)\right)|0\rangle
,\qquad x_i\in{\bf R}^s \eqno (\rm A.\theequation)$$

\vspace{.4cm}
\noindent satisfy the following orthogonality relation
\vspace{.4cm}
\setcounter{equation}{4}
$$\langle x_1,\ldots ,x_n|x'_1,\ldots ,x'_m\rangle
=\delta_{nm}\sum\limits_{\sigma}\prod\limits_{i=1}^{n}\delta(x_i-
x'_{\sigma(i)}),\eqno (\rm A.\theequation)$$

\vspace{.4cm}
\noindent where $\sigma$ is a permutation of the set
$\{1,\ldots ,n\}$, and completeness relation
\vspace{.4cm}
\setcounter{equation}{5}
$$\sum\limits_{n}\frac{1}{n!}\int d^sx_1\ldots
d^sx_n|x_1,\ldots ,x_n\rangle \langle x_1,\ldots ,x_n|=
I. \eqno (\rm A.\theequation)$$

\vspace{.4cm}
\noindent The vectors $|x_1,\ldots ,x_n\rangle$ form the
basis of the coordinate representation. The Bose operators
act on the basis vectors as follows
\vspace{.4cm}
\setcounter{equation}{6}
$$a(x)|x_1,\ldots ,x_n\rangle =
\sum\limits_{i=1}^{n}\delta(x-x_i)|x_1,\ldots ,\check
x_i,\ldots ,x_n\rangle ,$$
\hfill (A.6)
$$a^\dagger(x)|x_1,\ldots ,x_n\rangle =|x_1,\ldots
,x_n,x\rangle ,$$

\vspace{.4cm}
\noindent where the reversed hat over $x_i$ denotes that
this variable should be omitted from the set $\{x_1,\ldots
,x_n\}$.

\vspace{1.2cm}
\noindent {\bf Appendix B.~Functional coherent states
representation}

\vspace{.8cm}
We now recall the basic properties of the functional
coherent states. Consider the functional coherent
states $|u\rangle $, where $u\in L^2({\bf R}^s,d^sx)$ (the
complex Hilbert space of square-integrable functions). The
functional coherent states can be defined as the
eigenvectors of the Bose annihilation operators
\vspace{.4cm}
\setcounter{equation}{1}
$$a(x)|u\rangle =u(x)|u\rangle.  \eqno (\rm B.\theequation)$$

\vspace{.4cm}
\noindent The normalized functional coherent states can be
defined as
\vspace{.4cm}
\setcounter{equation}{2}
$$|u\rangle =\exp\left(-\frac{1}{2}\int
d^sx|u|^2\right)\exp\left(\int
d^sxu(x)a^\dagger(x)\right)|0\rangle. \eqno (\rm B.\theequation)$$

\vspace{.4cm}
\noindent These states are not orthogonal. We find
\vspace{.4cm}
\setcounter{equation}{3}
$$\langle u|v\rangle =\exp\left(-\frac{1}{2}\int
d^sx(|u|^2+|v|^2-2u^\ast v)\right). \eqno (\rm B.\theequation)$$

\vspace{.4cm}
\noindent The coherent states form the complete
(overcomplete) set. The formal resolution of the identity
can be written as
\vspace{.4cm}
\setcounter{equation}{4}
$$\int\limits_{\Omega^2}D^2u|u\rangle \langle u|=I, \eqno
(\rm B.\theequation)$$

\vspace{.4cm}
\noindent where $\Omega$ is the real space ${\cal D}'({\bf
R}^s)$ of Schwartz distributions or the real space ${\cal
S}'({\bf R}^s)$ of tempered distributions, $D^2u =
D(\hbox{Re}u)D(\hbox{Im}u)$ and the symbol $\exp\left(-\int
d^sxv^2\right)Dv$, where $v\in L^2_{\bf R}({\bf R}^s,d^sx)$
(the real Hilbert space of square-integrable functions)
designates the Gaussian measure on $\Omega$.

\vspace{.8cm}
\noindent The passage from the coordinate representation
to the functional coherent states representation is given by
\vspace{.4cm}
\setcounter{equation}{5}
$$\langle x_1,\ldots ,x_n|u\rangle =
\left(\prod\limits_{i=1}^{n}u(x_i)\right)\exp\left(-\frac{1}
{2}\int d^sx|u|^2\right).\eqno (\rm B.\theequation)$$

\vspace{.4cm}
\noindent Suppose now that we are given an arbitrary state
$|\phi\rangle $. It follows immediately from (A.5) and (B.5)
that the functional $\phi[u^\ast]=\langle u|\phi\rangle $ is
of the form
\vspace{.4cm}
\setcounter{equation}{6}
$$\phi[u^\ast]=\tilde \phi[u^\ast]\exp\left(-\frac{1}{2}\int
d^sx|u|^2\right), \eqno (\rm B.\theequation)$$

\vspace{.4cm}
\noindent where the functional $\tilde \phi[u^\ast]$ is
analytic.

\vspace{.8cm}
\noindent An easy calculation based on (B.1), (B.2) and (B.6)
shows that (B.6) can be written in the following abstract
basis independent form
\vspace{.4cm}
\setcounter{equation}{7}
$$|\phi\rangle =\tilde \phi[a^\dagger]|0\rangle.  \eqno (\rm B.\theequation)$$
\newpage
\noindent Taking into account (B.4) and (B.6) we find
\vspace{.4cm}
\setcounter{equation}{8}
$$\langle \phi|\psi\rangle
=\int\limits_{\Omega^2}D^2u\exp(-\smallint
d^sx|u|^2)\tilde \phi^\ast[u^\ast]\tilde \psi[u^\ast].
\eqno (\rm B.\theequation)$$

\vspace{.4cm}
\noindent The representation (B.8) is the functional
Bargmann representation. The Bose operators act in this
representation as follows
\vspace{.4cm}
$$a(x)\tilde \phi[u^\ast]=\frac{\delta}{\delta
u^\ast(x)}\tilde \phi[u^\ast],$$
\hfill (B.9)
$$a^\dagger(x)\tilde \phi[u^\ast]=u^\ast(x)\tilde
\phi[u^\ast].$$
\newpage
\noindent {\bf References}

\vspace{.8cm}
\noindent [1]\quad \parbox[t]{14cm}{A.C.~Newell, Solitons in Mathematics and
Physics (SIAM, Philadelphia, 1985).}\\

\noindent [2] \quad \parbox[t]{14cm}{K.~Kowalski, Physica
A {\bf 180} (1992) 156.}\\

\noindent [3] \quad \parbox[t]{14cm}{K.~Kowalski, Physica
A {\bf 145} (1987) 408.}\\

\noindent [4] \quad \parbox[t]{14cm}{K.~Kowalski, Physica
A {\bf 152} (1988) 98.}\\

\noindent [5] \quad \parbox[t]{14cm}{K.~Kowalski and
W.-H.~Steeb, Progr.~Theor.~Phys. {\bf 85} (1991) 713.}\\

\noindent [6] \quad \parbox[t]{14cm}{K.~Kowalski and
W.-H.~Steeb, Progr.~Theor.~Phys. {\bf 85} (1991) 975.}\\

\noindent [7] \quad \parbox[t]{14cm}{K.~Kowalski and
W.-H.~Steeb, Nonlinear Dynamical Systems and Carleman
Linearization (World Scientific, Singapore, 1991).}
\end{document}